# A newly developed multi-kilo-channel high-speed and precision waveform digitization system for neutrino experiments

H. Yang[a,b], T. Xue[a,b,1], L. Jiang[a,b], C. Xu, Q. Pan[a,b], B. Liang[a,b], G. Gong, B. Xu, Z. Wang, S. Chen, Y. Liu[a,b] and J. Li[a,b]

[a] *Key Laboratory of Particle & Radiation Imaging (Tsinghua University), Ministry of Education,*
*Qinghuayuan Street No. 1, Beijing, China*

[b] *Department of Engineering Physics, Tsinghua University,*
*Qinghuayuan Street No. 1, Beijing, China*
*E-mail:* xuetaothu@tsinghua.edu.cn

ABSTRACT: The Jinping Neutrino Experiment(JNE), conducted within the China Jinping Underground Laboratory, aims to detect and analyze of solar neutrinos, geo-neutrinos, and supernova neutrinos. A one-ton prototype will soon be in commision with an upgrade from 30 channels to 60 channels, which will increase the data bandwidth by one to two orders of magnitude and exceed the capacity of the current CAEN DAQ system. Additionally, enhancing the performance and flexibility of JNE DAQ system is crucial. This paper presents the design of a new Tsinghua DAQ system for the JNE and its performance and stability. The new Tsinghua DAQ(THDAQ) system for JNE is based on the cPCI protocol and demonstrates powerful performance improvements: ADC ENOB of the THDAQ system approximately exceeds 9.8-bit, marking a 14% improvement over the CAEN DAQ system; The maximum clock deviation within a single chassis is 85.6 ps, satisfying sub-nanosecond synchronization criteria; Each DAQ board features two QSFP+ optical ports with 82.5Gbps transmission capability, while the PCIe board supports a transmission rate of 100.2 Gbps. In addition, comparative experiments between the two systems were also tested in detail. The analysis results of waveform and charge spectrum prove the high stability of the THDAQ system. This provides a foundation for the 60-channel and 4000-channel DAQ systems.

---

[1] Corresponding author.

**Contents**

## 1. Introduction

The Jinping Neutrino Experiment (JNE) is located beneath Jinping Mountain at the China Jinping Underground Laboratory (CJPL). Shielded by a 2400-meter-thick rock overhead, the laboratory is ideal for neutrino detection. The scientific objective of JNE is to construct a 4000-channel liquid scintillator detector for the detection and analysis of solar neutrinos, geo-neutrinos, and supernova neutrinos. In 2017, JNE developed a 1-ton prototype for preliminary verification. Since its inception, JNE has achieved numerous academic achievements. The 1-ton prototype is comprised of 30 PMTs mounted on the exterior of the spherical detector. Outputs from each PMT are connected to CAEN DAQ system. JNE is planning an upgrade to the 1-ton prototype in the summer of 2024. The number of detector channels will increase from 30 to 60. The data bandwidth is set to expand by one to two orders of magnitude, surpassing the current capabilities of the CAEN DAQ systems. Furthermore, JNE aims to enhance the performance and flexibility of the DAQ system. Consequently, developing a new DAQ system for JNE is imperative. The objectives of the new Tsinghua DAQ(THDAQ) system are as follows:
- To increase the data bandwidth of the DAQ system, thereby detecting and analyzing a broader range of background and physical events.
- To enhance ADC performance, including ENOB and dynamic range, thereby improving the energy spectrum resolution and capturing higher-energy physical events.

- To enhance the DAQ systems scalability and development flexibility, enabling physicists to investigate a wider array of physical phenomena.
- To verify system stability, providing a foundation for the 60-channel and 4000-channel DAQ systems.

This paper outlines the THDAQ system architecture tailored for JNE in Section 2. Section 3 provides a comprehensive overview of the hardware design of the THDAQ system. Section 4 details the firmware implementation on the DAQ and PCIe boards. Section 5 presents the results from the hardware performance tests of the THDAQ system. Section 6 illustrates the comparative experimental results of the THDAQ system against the CAEN DAQ system. The conclusion and future directions are discussed in Section 7.

## 2. System architecture

**Figure 1** displays a picture of the JNE DAQ system. Through the splitter chassis, the signals of thirty PMTs is equally split into two channels received by THDAQ system and CAEN DAQ system. The THDAQ system operates in external trigger mode, receiving trigger signals from CAEN DAQ systems.

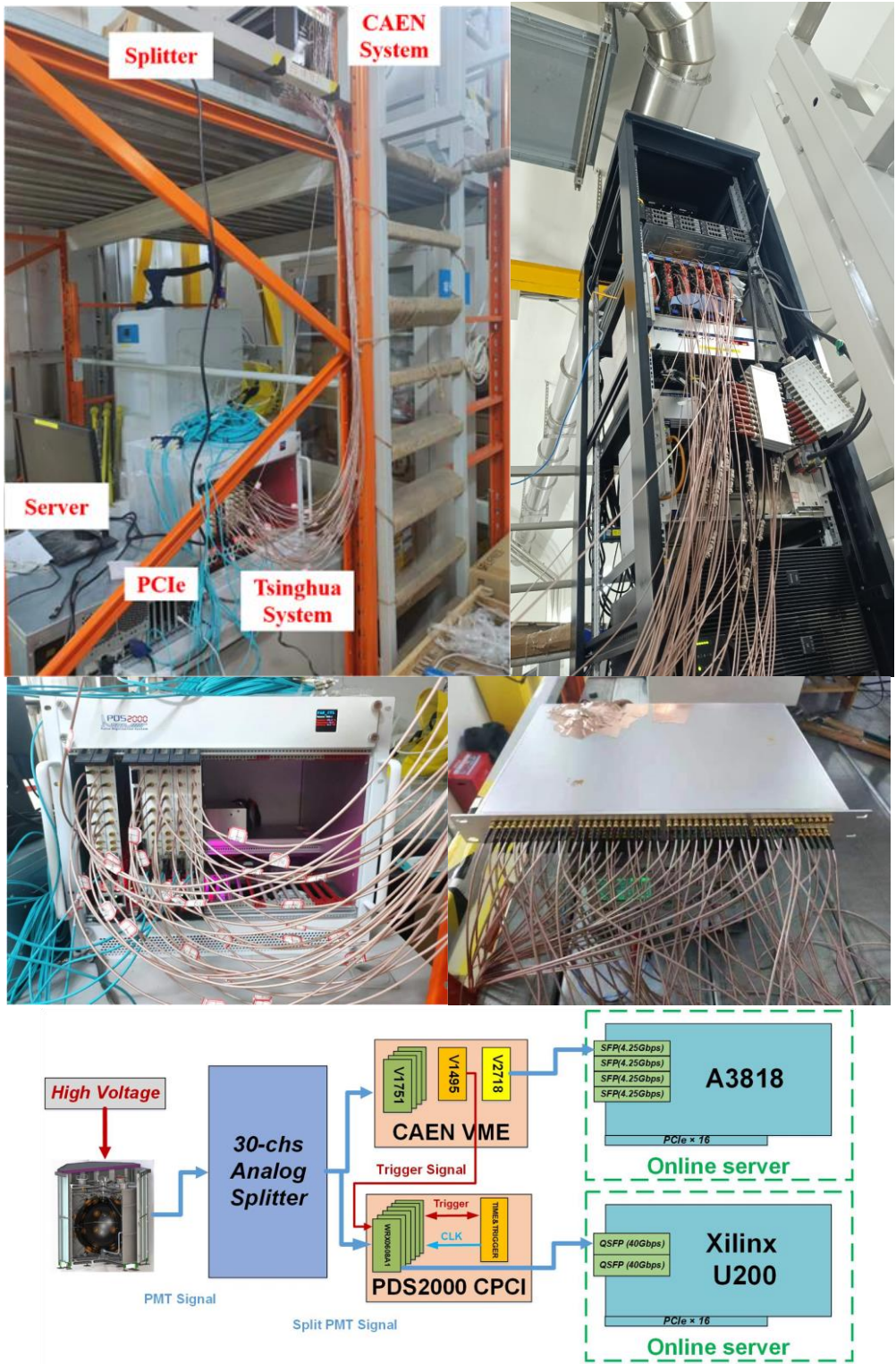

**Figure 1.** Experimental platform(top left), Caen DAQ system for JNE(top right), Tsinghua DAQ system for JNE(middle left), 30-channel Splitter chassis(middle right), Overall framework of experiment(bottom)

2.1 CAEN DAQ system

**Figure 2** shows block diagram of CAEN DAQ system for JNE. Data from all 30 channels are received by the CAEN V1495. A global trigger is generated when at least 25 channels exceed the threshold simultaneously. Following triggering, data from all channels are transferred to the CAEN V2718 board via the backplane. Ultimately, data is transmitted to the CAEN A2818 PCIe board by optical fiber and stored in ROOT format on the server.

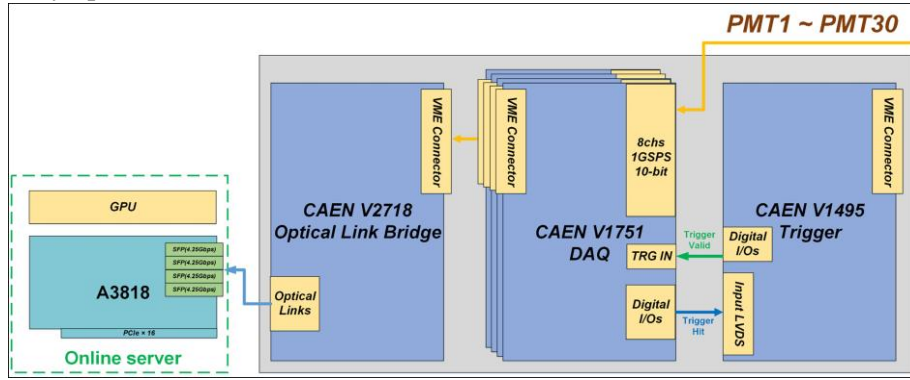

**Figure 2.** Block diagram of CAEN DAQ system for JNE

2.2 Tsinghua DAQ system

Within a standard 6U Compact PCI(cPCI) Express chassis, six DAQ boards, a trigger and clock distribution(TCD) board, and a 9-slot high-speed backplane are mounted. The received trigger signals are directly distributed to 6 DAQ boards by the TCD board. Each DAQ board is equipped with 6 channels of 13-bit/1GSPS ADC, which was self-developed by Tsinghua University. After receiving the trigger signal, the DAQ board packages and transmits the collected data to the PCIe board Xilinx U200 via optical fiber. Instructions are received by the PCIe board, which then sends data to the server by a 16-lane PCIe Gen3 link.

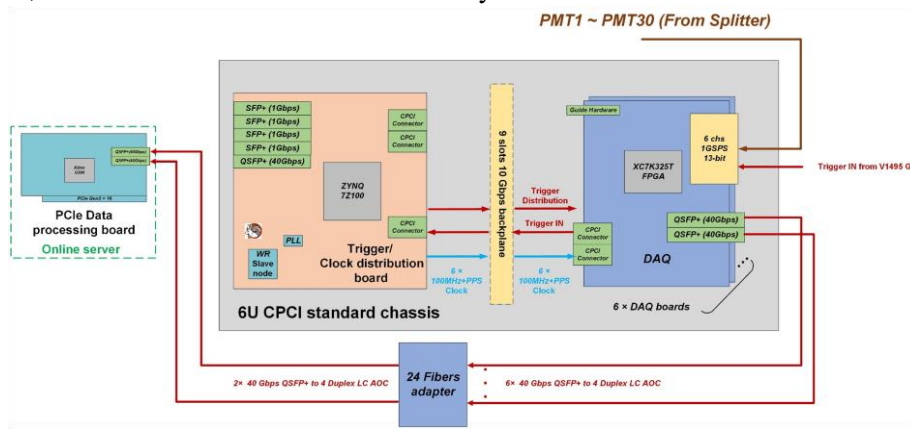

**Figure 3.** Block diagram of THDAQ system for JNE

## 3 Hardware design and implementation

### 3.1 Board of DAQ

The DAQ board and its functional block diagram are shown in **Figure 4**. The DAQ board, compliant with the 6U cPCI Express standard, features a 12-layer PCB structure utilizing FR4 material. Its architecture employs FPGA and ZYNQ, with the FPGA XC7K325T managing ADC data processing, digital shaping, online triggering, and high-speed data transmission, while the ZYNQ XC7Z010 primarily supervises board slow control, receives control parameters, and manages online firmware updates.

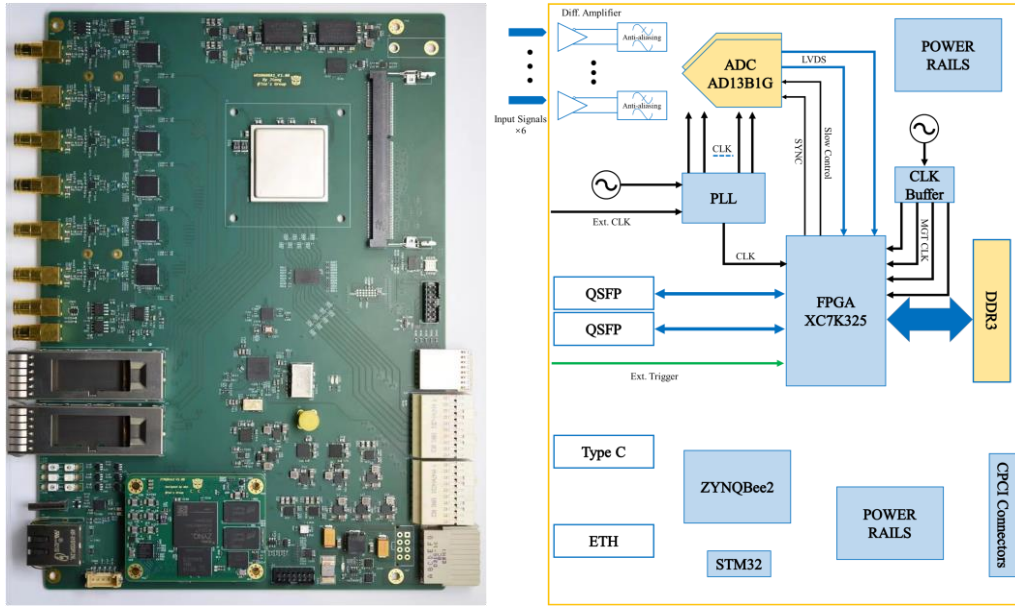

**Figure 4.** THDAQ board(left) and functional block diagram(right)

The clock of the DAQ board is provided by the dual cascade PLL LMK04803. The reference clock of the PLL is derived from a 100MHz synchronous clock at the backplane connector. In the dual cascade PLL structure of LMK04803, the voltage-controlled oscillator (VCO) of the first-stage PLL is an external ultra-low phase noise VCXO CVHD-950-100.000. The integrated phase noise of this VCXO is approximately 40 fs over the 12 kHz to 20 MHz range, with a voltage control sensitivity of ±25 ppm/V. Based on the external VCXO's characteristics, adjusting the first-stage PLL's loop bandwidth to 50 Hz effectively filters out high-frequency phase noise introduced by the reference clock and minimizes jitter. The voltage controlled oscillator of the second-stage PLL is an internal VCO with an adjustable frequency range of 1840~2030 MHz. The system locks the PLL frequency at 2000MHz, subsequently providing a 1GHz sampling clock to the ADC by frequency division.

The DAC can dynamically adjust each range of the ADC sampling channel to modify the circuit offset with a dynamic range of 1.3 V p-p. All ADCs digitize and transmit the waveform to the FPGA via the Low-Voltage Differential Signaling(LVDS). The FPGA then deserializes, recombines, and digitally filters the original data. This information, including waveform data, timestamp, and event sequence number, is packaged, merged, and subsequently transmitted to the PCIe board by optical fiber.

### 3.2 Board of Trigger and clock distribution

The TCD board is shown in **Figure 5**. This board, compliant with the 6U cPCI Express standard, comprises three main components: the motherboard, the ZYNQ module, and the CUTE-WR-A7 module. Its primary function is the distribution of global trigger and clock signals to all DAQ boards within the chassis. Trigger signals are distributed via multiple links connecting the ZYNQ to the backplane. These links originate from the ZYNQ pins, traverse the core board connector and backplane, and terminate at the FPGA pins on the DAQ boards, with all links being of equal length.

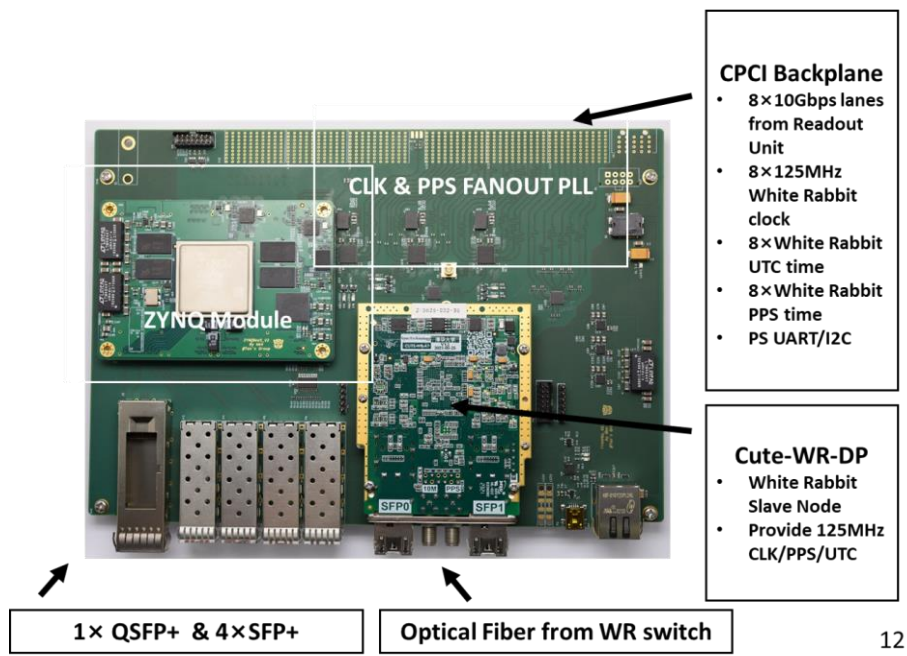

**Figure 5.** Trigger and clock distribution board

**Figure 6** shows the schematic diagram of the TCD board clock distribution. Synchronous clock generation employs two methods: either a 125MHz clock generated by the White Rabbit (WR) node or a 125MHz oscillator. The generated synchronous clock is divided into eight channels by the ADCLK948 chip and distributed to all DAQ boards on the same backplane within the chassis.

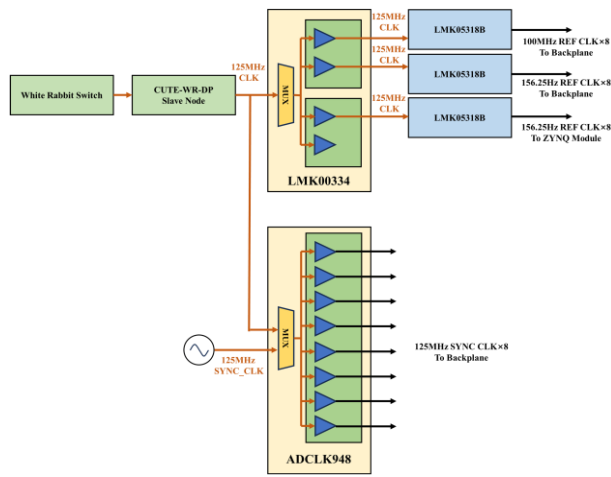

**Figure 6.** Schematic diagram of the TCD board clock distribution

### 3.3 Board of Backplane

The backplane serves as a carrier for high-speed data transmission and synchronous clock distribution. Furthermore, it supplies power to each slot and monitors board status information. As shown in **Figure 7**, the high-speed backplane is a 9-slot 6U cPCI Express model with a board thickness of 3.0mm and constructed from FR4 material. Each backplane accommodates one TCD board slot, along with eight DAQ board slots. A GTX high-speed serial link between the trigger and clock distribution board and each DAQ board, which can achieve data bandwidth up to 10.3125 Gbps. The TCD board distributes signals to each DAQ board: one 125 MHz synchronous clock, one PPS signal with TAI encoding, one 156.25 MHz serial transceiver reference clock, and seven LVDS links. It also distribute two I2C buses to collect power consumption and temperature data from each DAQ board.

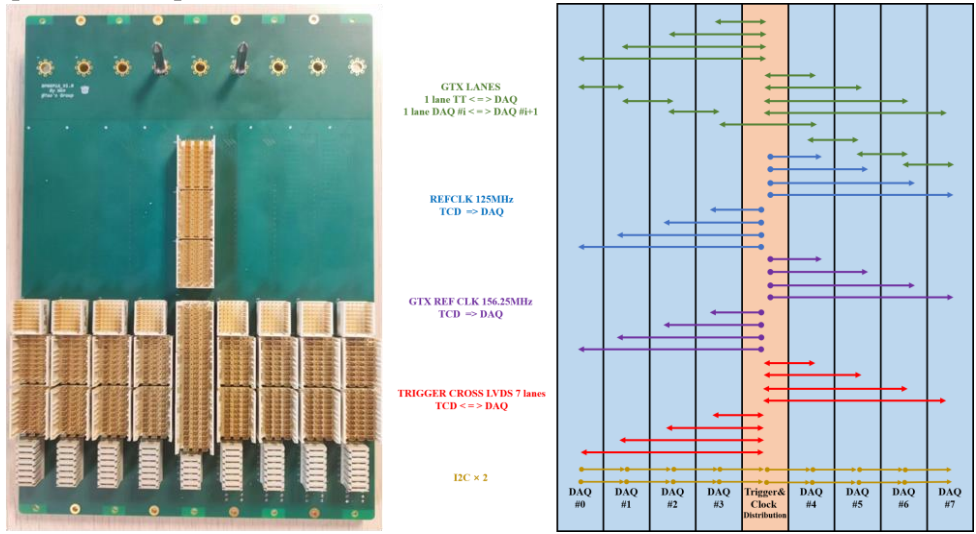

**Figure 7.** Backplane board(left) and signal interaction(right)

### 3.4 Compact PCI chassis

The cPCI chassis PDS2000 provides sixteen slots for DAQ boards, two for TCD boards, and three for slow control boards. A temperature and pressure monitoring and control system is integrated into the chassis, dynamically regulating fan speed in real-time. Situated in the upper right corner of the chassis, a Liquid Crystal Display (LCD) conveys critical information, including real-time temperature and air pressure readings. During extended full-power operation at typical temperature conditions, the internal temperature of the chassis is reliably maintained at approximately 23°C. The chassis power supply features the HRPG-1000-12 switching power supply, which is rated at 1000 W. Featuring the HRPG-1000-12 switching power supply rated at 1000 W, the chassis power supply distributes power through the backplane to the boards in each slot from this 12 V DC source. The chassis is shown in **Figure 8**.

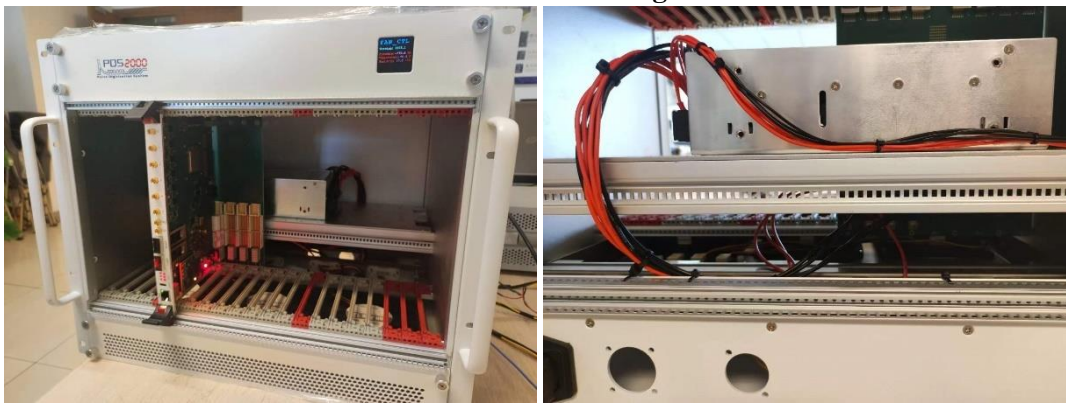

**Figure 8.** Compact PCI chassis PDS2000(left) and switching power HRPG-1000-12(right)

## 4 Firmware and software design

### 4.1 Firmware of DAQ Board

The firmware of the DAQ board mainly includes ADC data reorganization, ring buffer pool storage, synchronization timestamp generation, data packaging and data transmission, as shown in **Figure 9**. Due to interference from noise, only five ADC channels are utilized, with the sixth channel being unused. Each of the five ADC generates 13 pairs of LVDS data and a 500 MHz data clock output pair in DDR mode.

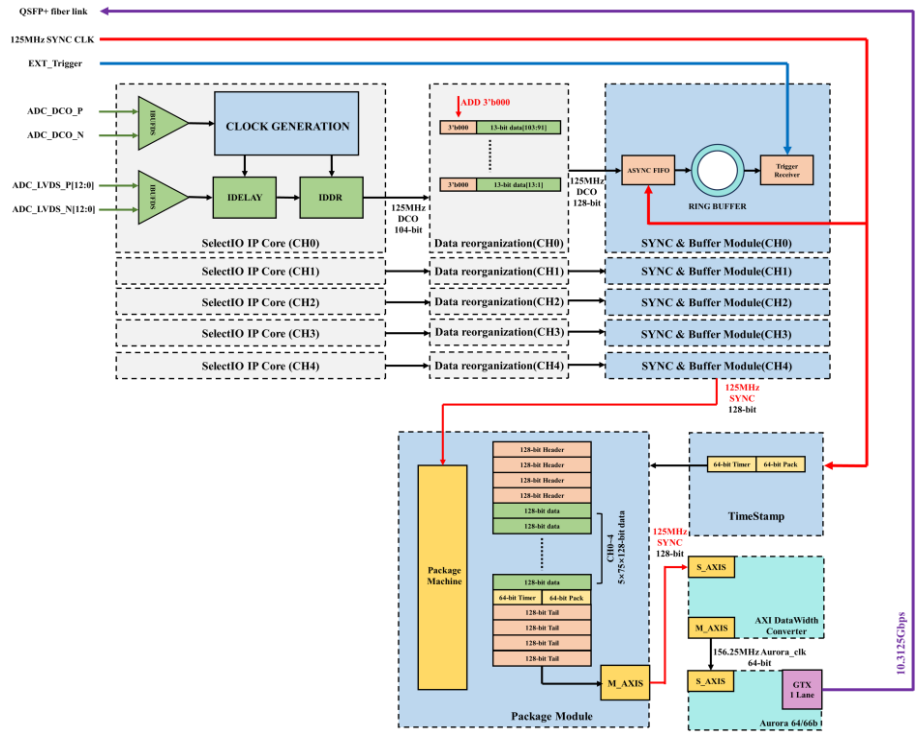

**Figure 9.** Firmware of the DAQ board

Firstly, the SelectIO IP receives the 500MHz clock and LVDS data from each ADC, with a serialization factor of 8. This IP generates a 125MHz data clock and 104-bit data combined from eight 13-bit data. Secondly, a 3-bit data sequence, 000, is added to the end of each 13-bit data within the 104-bit data to form 128-bit new data. Thirdly, 128-bit synchronous data is generated from 128-bit new data through Asynchronous FIFO IP and cached in the ring buffer. Finally, after receiving the trigger signal, the packaging module accesses seventy-five 128-bit data from each ring buffer. Each 128-bit data corresponds to an 8ns waveform, with a total of five 600ns waveforms being read. The complete event packet, cached in RAM, comprises a packet header, waveform data, packet tail, trigger number, and timestamp. Transmission of the event packet occurs through the Aurora IP core using 64b/66b encoding, achieving a bandwidth of 10.3125 Gbps.

### 4.2 Firmware of PCIe Board and software

The primary function of the PCIe board involves receiving data from six DAQ boards and transmitting this data to the server. Firstly, two Aurora IP cores receive the packaged data from six DAQ boards, each supporting a bandwidth of 10.3125 Gbps. Secondly, the AXIS Combiner

– 8 –

IP and AXI Datawidth Converter IP convert the data width to 512 bits, followed by input into an asynchronous FIFO. Next, the AXI Datamover IP and AXI Interconnect IP cache the data to onboard DDR4 memory. The DDR4 memory boasts a 16 GB capacity, a 64-bit width, and a read/write bandwidth of 2666 MT/s. Finally, data transfer from the PCIe board to the server via the PCIe 16-lane Gen3 link. At the core of the firmware lies the XDMA IP core from Xilinx, supporting the AXI4 bus protocol and offering an AXI4-Lite interface for reading and writing configuration data.

The XDMA driver is the basis for software function implementation. After loading the XDMA driver, the user program initiates a data transfer request. This request is then converted into an AXI-4 protocol-compliant instruction by the XDMA IP to access the data cached in DDR4. Control requests for customizable functions are transmitted via the AXI-LITE interface.

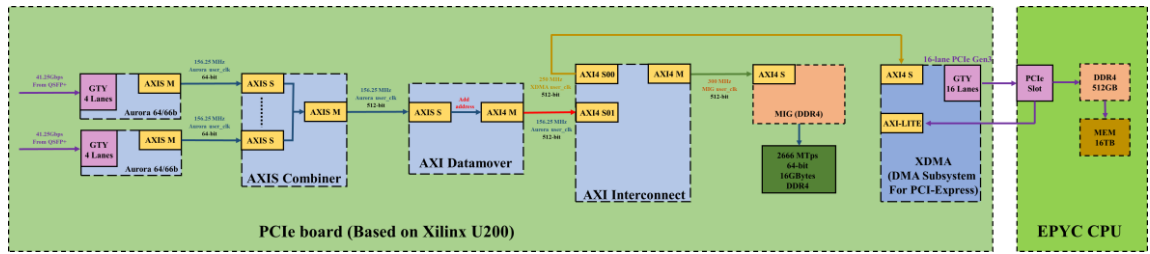

**Figure 10.** Firmware of PCIe board

## 5 Performance test

### 5.1 Analog performance test

The ENOB value of the Tsinghua ADC is measured using the sine wave fitting method, in accordance with the IEEE1241 standard. For this experiment, the sine wave generator SMB100A by Rohde & Schwarz is utilized. **Table 1** displays the ENOB values for thirty-six ADCs.

**Table 1.** ENOB value of each Tsinghua ADC

| Board | CH1 | CH2 | CH3 | CH4 | CH5 | CH6 |
|---|---|---|---|---|---|---|
| 0# | 9.71±0.16 | 9.91±0.15 | 10.09±0.09 | 9.88±0.14 | 9.93±0.07 | 9.36±0.05 |
| 1# | 9.66±0.23 | 9.88±0.19 | 10.08±0.11 | 9.89±0.19 | 9.89±0.09 | 9.33±0.06 |
| 2# | 9.57±0.25 | 9.90±0.22 | 9.99±0.29 | 9.86±0.20 | 9.90±0.11 | 8.94±0.06 |
| 3# | 9.67±0.19 | 9.92±0.15 | 10.09±0.08 | 9.91±0.16 | 9.93±0.08 | 9.33±0.04 |
| 4# | 9.05±0.28 | 9.93±0.18 | 10.10±0.12 | 9.91±0.20 | 9.92±0.11 | 9.29±0.06 |
| 5# | 9.57±0.16 | 9.88±0.15 | 10.07±0.07 | 9.87±0.14 | 9.84±0.06 | 9.26±0.04 |

The ENOB values of the sixth channel ADC on each DAQ board are significantly lower than those for the other five channels. This phenomenon is due to unreasonable PCB design. Furthermore, the typical ENOB values of Tsinghua ADCs range from 9.8 to 10.0, representing a 14% improvement over the ENOB values of the CAEN V1751 ADC which is 8.6.

## 5.2 Clock synchronization test

Clock synchronization is very important for DAQ systems. The experiment measure the 125MHz Test CLK outputs from various LMK04803 PLLs within the same chassis, utilizing the LeCroy WavePro 404HD oscilloscope, as shown in the **Figure 11**. The Test CLK on the DAQ board is connected to the oscilloscope via two 1-meter-long coaxial cables. The Test CLK of first DAQ board is connected to channel 2 of the oscilloscope, and the Test CLK of subsequent seven DAQ boards is connected to channel 3 in turn.

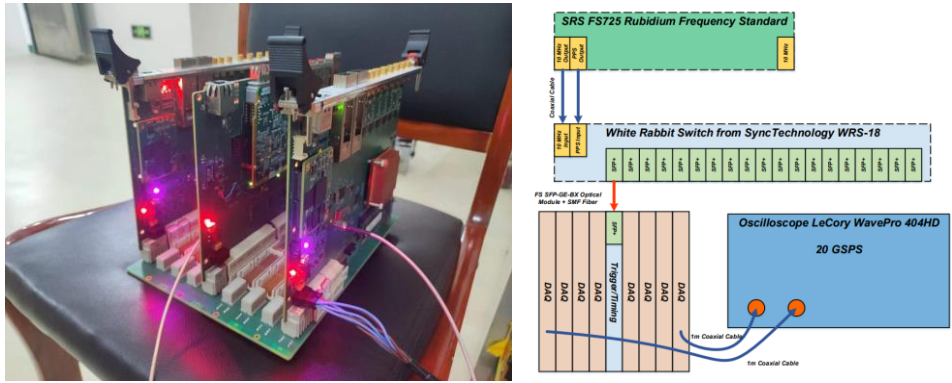

**Figure 11.** Picture(left) and block diagram of Clock synchronization

The measurement results reveal that the maximum clock deviation within a single chassis is 85.6 ps, satisfying the sub-nanosecond synchronization criteria of this DAQ system.

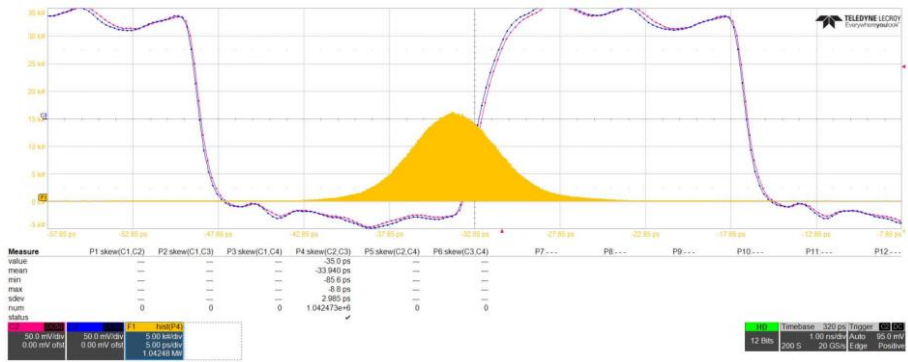

**Figure 12.** Clock deviation measurement results tested by oscilloscope

## 5.3 High-speed link test

A typical QSFP+ high-speed link eye diagram test result is shown in **Figure 13**. For all eight links, the Open UI exceeds 50%, and the bit error rate(BER) falls below $10^{-13}$. These test results confirm that the QSFP+ high-speed links signal integrity complies with the Xilinx Aurora 64b/66b transport standard. Figure displays the test results for the Aurora 64b/66b transmission. The cumulative transmitted number of data reached 0x1cd5df6edf7 without error, and the BER remained below $10^{-13}$.

– 10 –

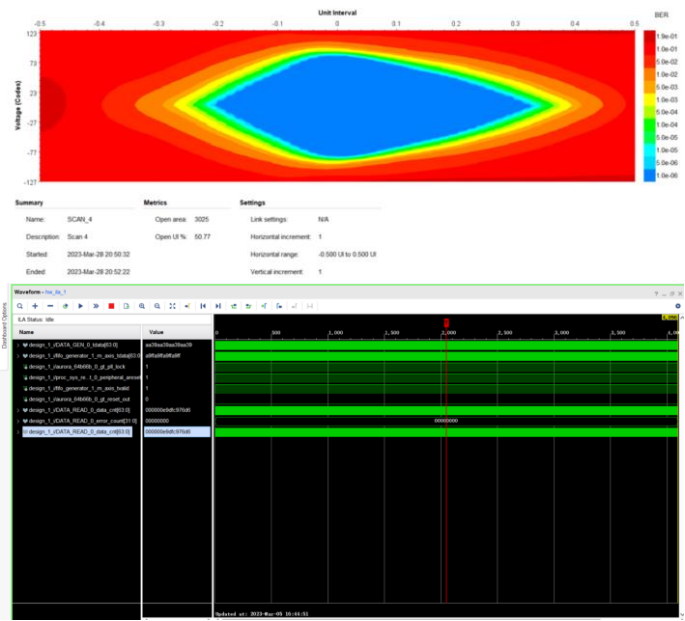

**Figure 13.** QSFP+ high-speed link eye diagram(top) and Aurora 64b/66b transmission test result(bottom)

Furthermore, the bandwidth of the 16-lane PCIe Gen3 is also tested. The average bandwidth of 10,000 transfers, each with a packet size of 512 MB, is 100.2 Gbps.

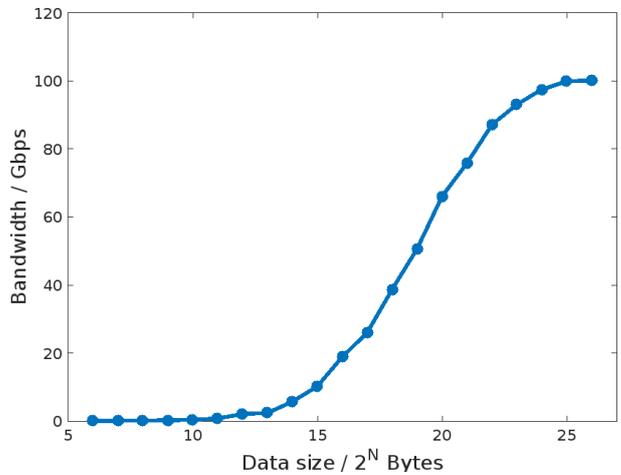

**Figure 14.** Average of bandwidth of 16-lane PCIe Gen3 in 10000 transfers. Packet size of is from 64kB to 512MB

## 6 1-ton prototype experimental results

To comprehensively compare the performance of two data readout systems and validate the stability of the THDAQ system, experiments were conducted including ADC baseline jitter, waveform, charge spectrum, and photo electron comparisons. Two DAQ systems simultaneously collected 408,677 events.

**Figure 15** shows the comparison results of baseline jitter standard deviation. It can be seen that the baseline jitter of the THDAQ system is less than 0.75 mV, which is better than the CAEN system whose baseline jitter is greater than 0.75 mV. These test results corroborate the ENOB findings detailed in Section 5.1, indicating significant improvement in the performance of the Tsinghua ADC.

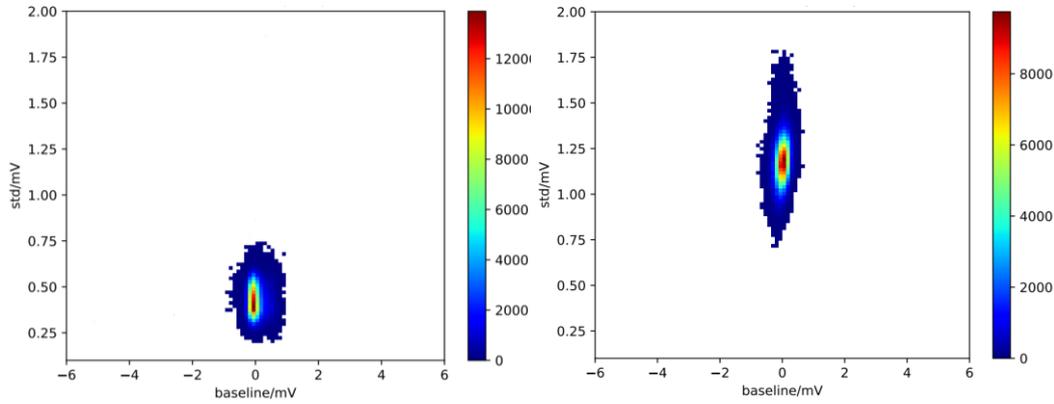

**Figure 15.** ADC baseline jitter results of THDAQ system(left) and CAEN DAQ system

Furthermore, to assess the data transmission quality of the Tsinghua system, waveforms collected by both systems were compared individually. The Tsinghua system recorded continuous trigger numbers, and the waveforms correspond directly one-to-one with those of the CAEN system, demonstrating no packet loss during data transmission. The average absolute difference of all waveform is less than 3mV. A typical waveform comparison result is shown in **Figure 16**.

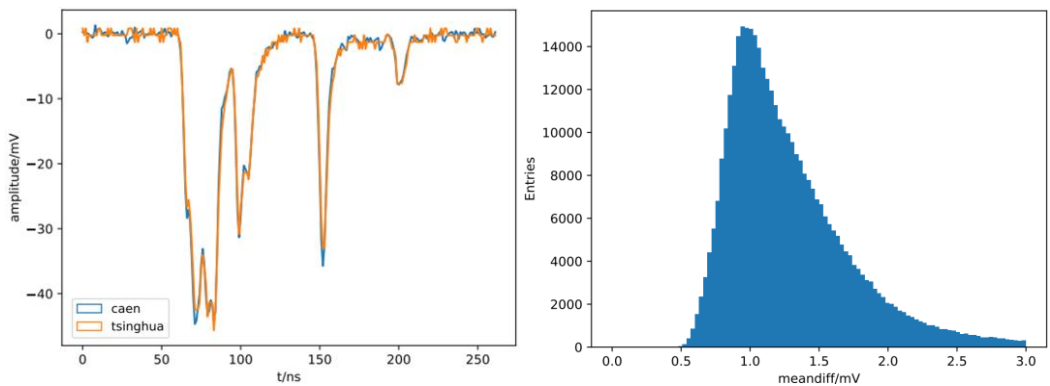

**Figure 16.** A typical waveform comparison result(left) and average absolute difference of the waveform(right)

The charge spectrum and PE spectrum of thirty PMTs were also compared to verify the accuracy of the waveforms collected by the THDAQ system. The charge spectrum collected by the Tsinghua system closely aligns with that of the CAEN system. Comparative analysis of the charge spectrum differences across the 30 channels revealed minor discrepancies between the two systems. Typical charge spectrum differences and PE spectrum are shown in **Figure 17**.

– 12 –

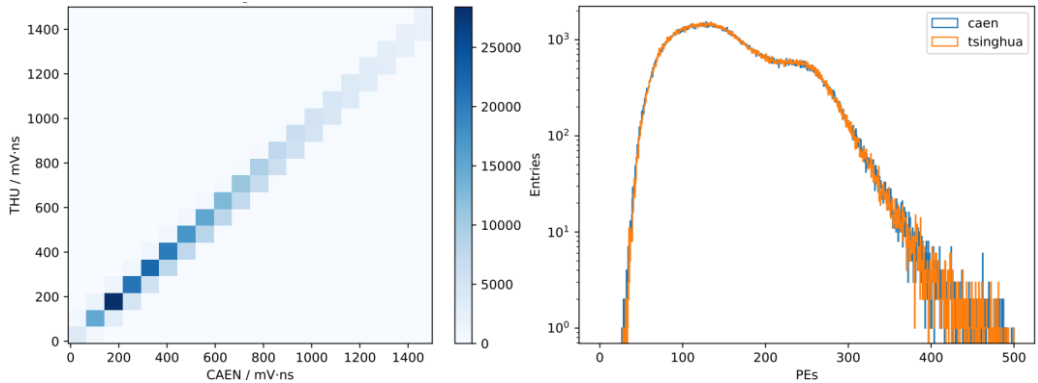

**Figure 17.** Charge spectrum(left) and PE spectrum(right) comparison between THDAQ system and CAEN DAQ system

## 7 Conclusion

This study details the design and evaluation of a 30-channel data acquisition system for the JNE. The THDAQ system comprises six DAQ boards, a TCD board, a Backplane board, a cPCI chassis, and a PCIe board. Five 13-bit 1GSPS ADC of each DAQ boards receive analog signals, which are then fed into the K7 FPGA for caching. After receiving a global trigger signal, each FPGA packages the cached data and transmits it to the PCIe board through optical fiber at a rate of 10.3125Gbps. The PCIe board receives data from the six DAQ boards, which is then cached in on-board DDR4 memory. Two 512 MB memory spaces are allocated on the board for ping-pong reading and writing, with the entire process managed by the host computer.

Additionally, this study comprehensively tests the hardware performance of the THDAQ system: the Tsinghua ADC ENOB approximately exceeds 9.8-bit, marking a 14% improvement over the CAEN system; The QSFP link eye diagram test results meet specified requirements, and the maximum clock skew within the chassis is 85.6ps; Furthermore, long-term testing alongside the CAEN system demonstrates Tsinghua system stability over extended periods and significant performance enhancements. The results show that the system can run stably for a long time and the performance improvement is obvious. This provides the basis for the implementation of complex algorithms and the development of a 4000-channel system.

The dual FMC carrier board utilizing the VPX protocol has been developed successful. Plans are to upgrade the hardware to a 60-channel data acquisition system based on KU060 and AD9695. PCIe boards based on KU060 are also under development. Additionally, previous studies have implemented the FPGA-GPU P2P method, which will facilitate the implementation of advanced triggering algorithms.

## Acknowledgments

We are grateful for the expertise and patient help from Yu Xue and Jianfeng Zhang, technical engineers in the electronics workshop at DEP.